\documentclass[traditabstract]{aa} 
\usepackage{txfonts,psfig}
\usepackage{color}

\begin{document}
\title{Making the observational parsimonious richness a working mass proxy}
\titlerunning{} 
\author{S. Andreon\inst{1}}
\authorrunning{S. Andreon}
\institute{
INAF--Osservatorio Astronomico di Brera, via Brera 28, 20121, Milano, Italy,
\email{stefano.andreon@brera.inaf.it} \\
}
\date{Accepted ... Received ...}
\abstract{Richness, i.e., the number of bright cluster galaxies, is known to  
correlate with the cluster mass, however, to exploit it as mass proxy 
we need a way to estimate the aperture in which galaxies should be counted
that minimizes the scatter between mass and richness. 
In this work, 
using a sample of 39 clusters with accurate
caustic masses at $0.1<z<0.22$, we first show that the scatter between 
mass and richness derived from survey data is
negligibly small, as small as best 
mass proxies. The scatter turns out to be smaller
than in some previous works and has a 90\% upper limit 
of $0.05$ dex in mass.
The current sample, adjoining 76 additional
clusters analyzed in previous works, establishes
an almost scatterless, minimally evolving (if at all), 
mass-richness scaling in the redshift range $0.03<z<0.55$.
We then exploit this negligible scatter to derive 
the reference aperture to be used to compute richness and to predict the mass 
of cluster samples. These predicted masses have a total
$0.16$ dex scatter with caustic mass, about half of which is not
intrinsic to the proxy, but related to the noisiness of the 
caustic masses used for test proxy performances.
These results make
richness-based masses of best quality and
available for large samples 
at a low observational cost.
}

\maketitle

smaller than belived before

\section{Introduction}

In the next few years, numerous optical and near-infrared
surveys, e.g., the Panoramic
Survey Telescope and Rapid Response Systems (Kaiser et al. 2010),
the Dark Energy Survey (Abbott et al. 2005), 
Large Synoptic Survey Telescope (Ivezic et al. 2008),
Hyper-Suprime Camera (Takada 2010), and Euclid (Laureijs et al. 2011) 
are expected to generate
galaxy catalogs over several thousands of square degrees
to sufficient depth to reliably detect galaxy clusters up to $z \sim 2$.
Indeed, the most distant cluster, JKCS\,041 at $z=1.803$ (Andreon
et al. 2009, 2014; Newman et al. 2014) has been discovered on the
UKIRT infrared deep sky survey (Lawrence et al. 2007) and
Euclid will reach deeper magnitudes over one thousand times 
wider fields (Laureijs et al. 2011).

One of the main goals of these surveys
is to probe the expansion history of the Universe and
place tight constraints on cosmological parameters using
galaxy clusters. For example,
accurate measurements of cluster abundance as a function
of cluster mass and redshift can provide important
constraints on cosmological parameters such as $\sigma_8$ 
and $w$ (see, e.g., Vikhlinin et al. 2009).
However, the mass of a galaxy cluster is not directly measurable.
These surveys will therefore rely on mass proxies
and scaling relations
between these observables and mass. Calibration of mass-observable
scaling relations is therefore currently a high-priority observational
goal because the use of galaxy clusters as cosmological probes is
currently limited by our ability to relate observable properties
and cluster mass through so-called observable-mass relations.
For example, the major source of uncertainty 
in cosmological estimates by South Pole Telescope clusters is the currently
available mass-proxy calibration, not sample size (Reichardt 
et al. 2013).

For photometric surveys, optical richness is the primary mass proxy,
although stellar mass (i.e., the total cluster luminosity) has also been
proposed as accurate alternative (Andreon 2012). The former has
a scatter of 0.18 dex with mass (Andreon \& Hurn 2010,  
revised downward in Sec.~3.1), the latter has 90 \%
upper limit scatter of  $0.06$ dex. These performances are however
those achieved with knowledge of the radius in which galaxies
have to be counted and are expected to degrade when this information
is not available, i.e., when richness is used as a mass proxy. 
Indeed, richness has a $0.29$ dex scatter with mass when $r_{200}$ is
inferred from photometric data (Andreon \& Hurn 2010, 
revised downward in Sec.~3.2),which is much worse than if
$r_{200}$ were known. Observables with large scatter 
are particularly problematic because of
the sensitivity of cluster abundance on the
uncertainty in the scatter of the mass-observable relation 
(Lima \& Hu 2005), and the latter are expected to
increase with scatter. Consequently,  
a richness estimator that
minimizes the scatter in the richness-mass relation is highly
desirable to minimize the dilution of the cosmological information.

Therefore, an ideal mass proxy should be
characterized by a low intrinsic scatter with cluster mass.
It should also satisfy two more requirements.
First, the proxy should be observationally parsimonious to
obtain: low-scatter mass proxies are of little
utility if they are unavailable 
because they require the acquisition of data 
challenging to obtain or their measurement is unfeasible.
Second, the proxy should be relatively insensitive to the cluster 
dynamical state: a proxy that heavily relies on
hydrostatic or dynamical equilibrium, as some X-ray mass observables,
is of low utility for most of the clusters.

Once obtained, mass proxies need to be calibrated.
Gravitational lensing offers a way to calibrate the richness-mass
relation with the key advantage that the derived
cluster mass does not rely on the assumption of hydrostatic
or dynamical equilibrium. 
However, 
converting 
the lensing observable (reduced shear) into mass is challenging,
as is nicely illustrated in detail by
Hoekstra et al. (2015) and also is summarized in Andreon (2015b).
Furthermore,
the weak-lensing signal is also sensitive to mass in other
structures ``close" to the cluster (within say 1000 Mpc, because
the weak-lensing signal changes little on these scales) 
such as large-scale-structure. By measuring
the total mass projected along a long line
of sight, 
lensing masses have an unavoidable minimal uncertainty of
20 to 50\% (Meneghetti et al. 2010; Becker \& Kravtsov 2011) because of 
cluster asphericity and large-scale structure.

Caustics (Diaferio \& Geller 1997; Diaferio 1999) are also observationally
demanding, because caustics rely on measurements
of the escape velocity of galaxies, and are not 
affected by the cluster non-equilibrium, as weak lensing.
Furthermore, and in contrast to the latter, caustics
are not affected
by the correlated large-scale structures
along the line of sight (Diaferio 1999; Serra et al. 2011;
Geller et al. 2013) because 
removing background/foreground at $\Delta v > 3000$ km/s
is straightforward with spectroscopy that is needed anyway for caustics.
This is magnificently illustrated by the cluster pair 
MS0906.5+1110/Abell 750: these clusters are 
offset by only 3000 km/s and are almost
on the same line of sight. Caustics easily distinguish the two
objects and two entries are found in Rines et al. (2013) mass catalog.
Instead, lensing does not distinguish the two clusters and one single entry is found in
Hoekstra et al. (2015) catalog, with mass given by the sum of the two
clusters.

In this paper, we use an X-ray selected 
sample of 39 clusters with accurate
caustic masses at $0.1<z<0.22$ to improve upon our past mass-richness
calibration (Andreon \& Hurn 2010) 
and we show that the scatter between 
mass and richness derived from survey data is
negligibly small, and indeed smaller than quoted in previous works. 
Our current analysis benefits of an improved understanding
of caustic errors (Serra et al. 2011). We then exploit 
this negligible scatter to derive 
the reference aperture to be used to compute richness and to predict the mass 
of cluster samples, outperforming our (and other people) previous works.
Finally, we compare the performances of richness
and of a much advertised low-scatter mass proxy, the pseudo-pressure 
$Y_{SZ}$.  The purpose of this work is not to find the optimal
overdensity $\Delta$, which minimizes the scatter between predicted
and true mass with a variable $\Delta$. Instead, we assumes that $M_{\Delta=200}$
is the target mass, and look at minimizing the scatter with this
mass at this fixed overdensity.

Throughout this paper, we assume $\Omega_M=0.3$, $\Omega_\Lambda=0.7$, 
and $H_0=70$ km s$^{-1}$ Mpc$^{-1}$. Magnitudes are in the AB system.
We use the 2003 version of Bruzual \& Charlot (2003, hereafter BC03) stellar 
population synthesis
models with solar metallicity, a Salpeter initial 
mass function (IMF) and a $z_f=3$.
Results of stochastic computations are given
in the form $x\pm y$, where $x$ and $y$ are 
the posterior mean and standard deviation. The latter also
corresponds to 68 \% intervals because we only summarized
posteriors close to Gaussian in this way. All logarithms are in base 10.

\section{Samples and data} 

\subsection{The cluster sample}

Our starting point is the HECS (Hectospec Cluster Survey, 
Rines et al. 2013) cluster catalog.
Clusters are: a) X-ray flux-selected; 
b) with $0.1<z<0.3$ (to allow a good caustic
measurement); and c) in the SDSS DR6 footprint.
The cluster catalog lists the cluster center, 
radius $r_{200}$, and mass within $r_{200}$,
$M_{200}$, derived by the caustic technique\footnote{$r_\Delta$ is the
radius within which the enclosed average mass density is $\Delta$
times the critical density at the cluster redshift.} (Diaferio \& Geller 1997; Diaferio 1999).
These masses have, as mentioned, the advantage of not assuming 
the hydrostatic equilibrium 
or the relaxed status of the cluster and are
computed using, on average, 177 member galaxies per cluster.
These masses are three-dimensional (i.e., refer to the mass inside
the sphere of radius $r_{200}$) and are derived assuming a
spherical geometry. On the other
hand, and similar to other mass estimates, 
the computation of their uncertainty is
challenging (Serra et al. 2011; Gifford et al. 2013).
Because of the finite sampling of the velocity field, these
masses have a $\sim20$\% error (Serra et al. 2011). We adopt
this figure, instead of the noisier and suspect (too small) values 
listed in Rines et al. (2013),
and we account for the lack of precise knowledge of the true
error, allowing the latter
to be 95\% of the times 
within a factor of 2 of the value derived from simulations (20\%)
and marginalizing
over this source of uncertainty (see Andreon 
\& Hurn 2010 concerning how to deal with noisy errors).

From the HECS cluster catalog we
only keep clusters whose (passively evolved) limiting magnitude 
$M_{V,z=0}=-20 $ mag is
brighter than $r=20$ mag, a conservative completeness value 
of the SDSS (Ivezic et al. 2002).
This reduced the sample to clusters with $0.10<z\lesssim 0.22$, 
where the precise upper end depends on
the value of the Galactic extinction in the cluster direction.
From this sample, we only removed three clusters:
a) the cluster pair MS0906.5+1110/Abell 750 because 
the richness of these clusters
cannot be easily derived from photometry alone 
(see Figure~11 in Geller et al.  2013) because they are
almost on the same line of sight and separated by
only 3000 km/s; and b) the bimodal Abell 2055 cluster
because it is formed by two clumps apart by about
$r_{200}$. The spherical symmetry implicit in
the caustic method is certainly not satisfied by this cluster, at least
at the $r_{200}$ scale.
Table 1 lists the 39 studied clusters, whose distribution in the redshift-mass 
plane is depicted in Figure~1.

\begin{figure}
\centerline{
\psfig{figure=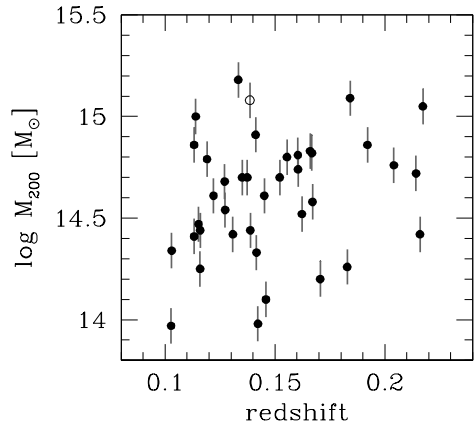,width=7truecm,clip=}
}
\caption[h]{Mass vs redshift plot of the studied cluster sample. 
The open point is Abell 1068. 
}
\end{figure}

\begin{table*}
\caption{Cluster sample and results of the analysis. 
The table lists cluster id, redshift $z$, mass $M_{200}$ from Rines et al. (2012), followed
by values derived in this work: richness $n_{200}$,
observed galaxy counts in the cluster and control field directions ($obstot$ and $obsbkg$), their solid angle ratios
$C$. Columns eight and nine give the predicted mass $\widehat{M_{200}}$ and richness $\widehat{n_{200}}$ 
we derived only using richness. 
Masses are within spheres, richnesses within cylinders. Masses in column (3)
have 0.08 dex errors, those in column (8) have a 0.16 dex total error. 
Cluster coordinates are in Rines et al. (2012).
}
\footnotesize
\begin{tabular}{l r r r r r r r r}
\hline
Cluster id & $z$ & $\log M_{200}$ & $n_{200}$ & $obstot$ & $obsbkg$ & $C$ & $\widehat{\log M_{200}}$ & $\widehat{n_{200}}$ \\ 
 & & $[M_\odot]$ & & & & & $[M_\odot]$ &\\
(1) & (2) & (3) & (4) & (5) & (6) & (7) & (8) & (9) \\
\hline
Zw1478 & $0.103$ & $13.97$ & $21\pm5$ & $24$ & $43$ & $16.940$  & $13.99$ & $21\pm5$ \\ 
A1235 & $0.103$ & $14.34$ & $51\pm8$ & $58$ & $68$ & $9.733$   &       $14.54$ & $57\pm8$ \\ 
A2034 & $0.113$ & $14.86$ & $102\pm11$ & $114$ & $68$ & $5.422$ & $14.89$ & $108\pm11$ \\ 
Zw8197 & $0.113$ & $14.41$ & $41\pm7$ & $47$ & $67$ & $10.760$  & $14.35$ & $40\pm7$ \\ 
A2069 & $0.114$ & $15.00$ & $111\pm13$ & $148$ & $164$ & $4.425$ &      $14.92$ & $110\pm12$ \\ 
A1302 & $0.115$ & $14.47$ & $52\pm8$ & $58$ & $60$ & $10.140$ & 	      $14.50$ & $53\pm8$ \\ 
A1361 & $0.116$ & $14.25$ & $25\pm6$ & $30$ & $74$ & $14.430$ & 	      $14.08$ & $25\pm5$ \\ 
A1366 & $0.116$ & $14.44$ & $51\pm8$ & $57$ & $60$ & $10.860$ & 	      $14.51$ & $54\pm8$ \\ 
A2050 & $0.119$ & $14.79$ & $71\pm9$ & $83$ & $77$ & $6.689$ &         $14.60$ & $63\pm9$ \\ 
A1033 & $0.122$ & $14.61$ & $59\pm8$ & $68$ & $86$ & $9.306$ &         $14.57$ & $59\pm8$ \\ 
A655 & $0.127$ & $14.68$ & $112\pm11$ & $120$ & $73$ & $9.113$ &	$15.03$ & $135\pm12$ \\ 
A646 & $0.127$ & $14.54$ & $47\pm7$ & $54$ & $83$ & $11.250$ &  	       $14.44$ & $48\pm7$ \\ 
A1930 & $0.131$ & $14.42$ & $29\pm6$ & $36$ & $100$ & $14.250$ &	  $14.01$ & $22\pm5$ \\ 
A1437 & $0.133$ & $15.18$ & $120\pm12$ & $144$ & $109$ & $4.614$ &	$14.91$ & $110\pm11$ \\ 
A1132 & $0.135$ & $14.70$ & $64\pm9$ & $76$ & $122$ & $9.948$ & 	       $14.59$ & $62\pm9$ \\ 
A795 & $0.137$ & $14.70$ & $89\pm10$ & $101$ & $123$ & $10.290$  &	$14.81$ & $92\pm10$ \\ 
A1068 & $0.139$ & $15.08$ & $50\pm9$ & $76$ & $151$ & $5.865$ & 	       $14.33$ & $39\pm7$ \\ 
A1918 & $0.139$ & $14.44$ & $45\pm7$ & $53$ & $122$ & $15.680$ &	 $14.39$ & $44\pm7$ \\ 
A1413 & $0.141$ & $14.91$ & $103\pm11$ & $120$ & $137$ & $7.865$ &	 $14.86$ & $100\pm11$ \\ 
A990 & $0.142$ & $14.33$ & $49\pm8$ & $60$ & $206$ & $19.220$ & 	       $14.50$ & $53\pm8$ \\ 
Zw3179 & $0.142$ & $13.98$ & $32\pm6$ & $36$ & $133$ & $33.320$ &	  $14.24$ & $33\pm6$ \\ 
A667 & $0.145$ & $14.61$ & $47\pm8$ & $57$ & $136$ & $13.200$ & 	       $14.38$ & $43\pm7$ \\ 
A1978 & $0.146$ & $14.10$ & $34\pm6$ & $40$ & $177$ & $29.230$ &	  $14.38$ & $43\pm7$ \\ 
A2009 & $0.152$ & $14.70$ & $81\pm10$ & $94$ & $161$ & $12.630$ &	$14.75$ & $82\pm10$ \\ 
A980 & $0.155$ & $14.80$ & $84\pm10$ & $101$ & $192$ & $11.220$ &	$14.73$ & $79\pm10$ \\ 
RXJ1720 & $0.160$ & $14.81$ & $74\pm10$ & $89$ & $179$ & $11.890$ &	 $14.66$ & $70\pm9$ \\ 
A2259 & $0.160$ & $14.74$ & $61\pm9$ & $75$ & $178$ & $13.180$ &	 $14.51$ & $54\pm8$ \\ 
A1902 & $0.162$ & $14.52$ & $69\pm9$ & $77$ & $142$ & $18.780$ &	 $14.71$ & $76\pm9$ \\ 
A1914 & $0.166$ & $14.83$ & $114\pm11$ & $126$ & $150$ & $12.160$ &    $14.98$ & $124\pm12$ \\ 
A1553 & $0.167$ & $14.82$ & $91\pm10$ & $104$ & $167$ & $12.610$ &	$14.81$ & $91\pm10$ \\ 
A1201 & $0.167$ & $14.58$ & $75\pm9$ & $89$ & $254$ & $18.180$ &	$14.76$ & $84\pm10$ \\ 
A1204 & $0.171$ & $14.20$ & $37\pm6$ & $41$ & $149$ & $33.860$ &	 $14.31$ & $38\pm7$ \\ 
A2187 & $0.183$ & $14.26$ & $40\pm7$ & $46$ & $225$ & $35.260$ &	  $14.38$ & $42\pm7$ \\ 
A1689 & $0.184$ & $15.09$ & $160\pm13$ & $180$ & $193$ & $9.919$ &	$15.13$ & $160\pm14$ \\ 
A1246 & $0.192$ & $14.86$ & $103\pm11$ & $121$ & $277$ & $15.240$ &    $14.88$ & $104\pm11$ \\ 
A963 & $0.204$ & $14.76$ & $75\pm9$ & $89$ & $282$ & $20.020$ & 	       $14.68$ & $73\pm9$ \\ 
A1423 & $0.214$ & $14.72$ & $83\pm10$ & $97$ & $321$ & $23.120$ &	 $14.78$ & $87\pm10$ \\ 
Zw2701 & $0.216$ & $14.42$ & $42\pm7$ & $48$ & $239$ & $37.400$ &	  $14.27$ & $35\pm6$ \\ 
A773 & $0.217$ & $15.05$ & $119\pm12$ & $148$ & $413$ & $14.330$ &	$14.93$ & $114\pm12$ \\ 
\hline      
\end{tabular} 
\end{table*}

\begin{figure*}
\centerline{
\psfig{figure=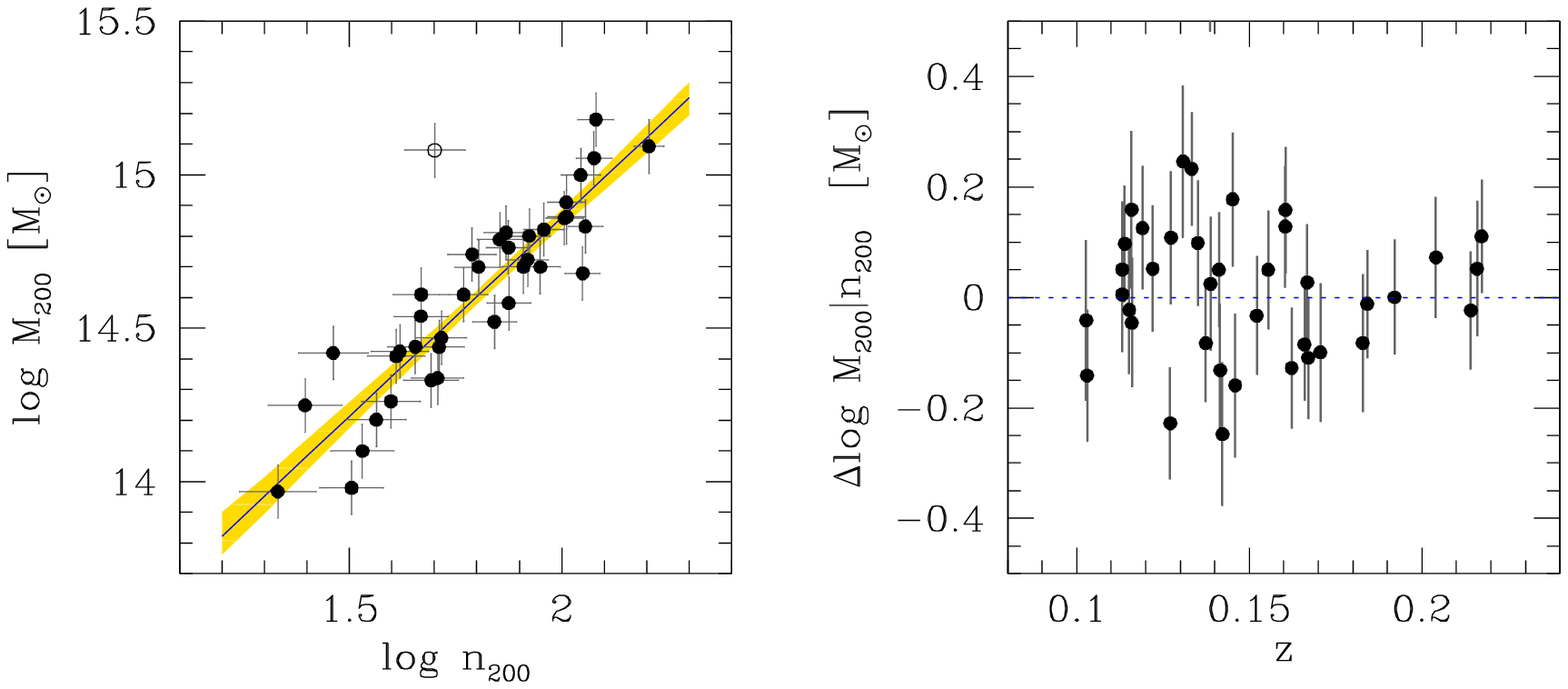,height=5truecm,clip=}
\psfig{figure=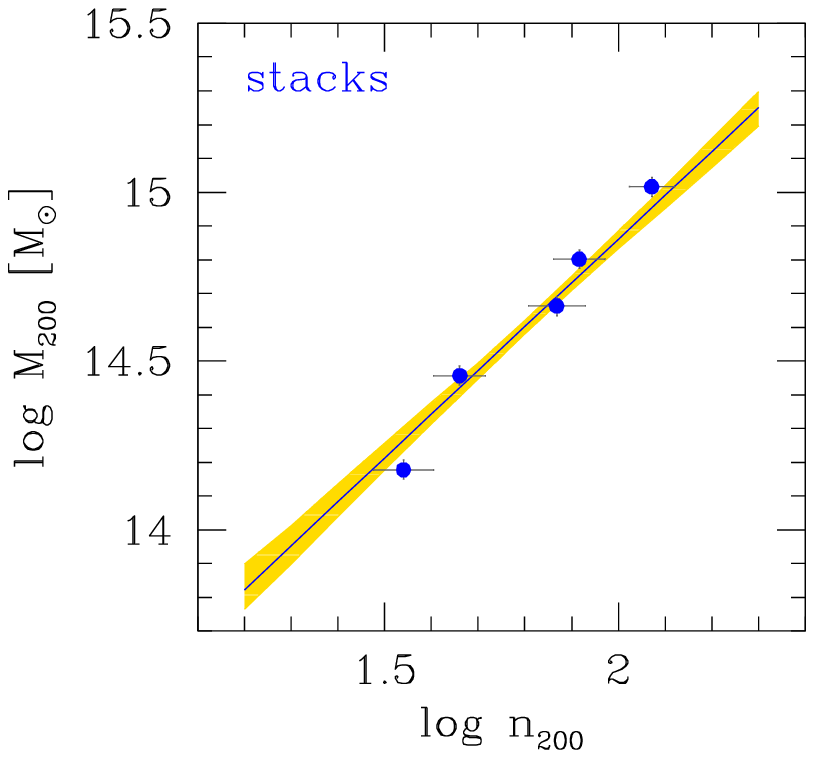,height=5truecm,clip=}
}
\caption[h]{
Richness-mass scaling (left-hand and right-hand panels) and residuals (observed minus
expected) as a function of redshift (central panel). 
The solid line marks the mean regression line 
(of $\log M_{200}$ on $\log n_{200}$)
fitted to the individual galaxy data, 
while the shading indicates the 68\% uncertainty
(highest posterior density interval). The right-hand panel 
combines clusters in stacks of eight
clusters each, with the exception of the most massive point, 
composed of just six clusters.
Masses are corrected for the negligible best-fit evolution.
Points and approximated
error bars are derived 
by adding errors summed in quadrature, as commonly done in the literature.
The open point (Abell 1068) is not fitted and is out of scale in the central panel.
}
\end{figure*}

\subsection{The data and derivation of cluster richness}

Our analysis closely follows 
previous works, in particular Andreon \& Hurn (2010),
which uses data from an earlier SDSS data release (for a different cluster
sample), and
Andreon \& Congdon (2014), which extends the analysis to account
for evolution of the mass-proxy scaling.

Basically, we aim to count red members within a specified luminosity range
and color, and within the $r_{200}$ radius, as already done for other
clusters (Andreon 2006a, 2008; Andreon et al. 2008; Andreon \& Hurn 2010,
Andreon \& Berg\'e 2012). 
For each cluster, we extracted the galaxy catalogs 
from the Sloan Digital Sky Survey (hereafter
SDSS) $12^{th}$ data release (Alam et al. 2015).
Total galaxy magnitudes refer to
``cmodel", while colors are based on ``model" magnitude.
Colors are corrected for
the color-magnitude slope, although this is a minor correction 
given the small magnitude range explored. 
We adopt a simple definition of `red', by only counting galaxies 
within $0.1$ redward and $0.2$ blueward in $g-r$ of the 
color-magnitude relation.

Some of the red galaxies in the cluster line of sight are actually in
the cluster fore/background. The contribution from background galaxies is
estimated, as usual, from a reference direction (e.g., Zwicky 1957; Oemler
1974; Andreon, Punzi \& Grado 2005). The reference direction is 
formed of three octants, free of contaminating structures (other clusters)
and not badly affected by the SDSS imaging masks, 
of a corona centered on the studied cluster with inner radius 3 Mpc and 
outer radius 1 degree, therefore fully guaranteeing 
homogeneous data for cluster and control field. 
The precise background estimation
is however of secondary importance: we found that if a single 
redshift-dependent value of 
the background counts  
per unit arcmin$^2$ were
used for all clusters, indistinguishable results would be obtained.

The derived (projected) richness values are listed in Table 1. 
Since richness is based on galaxy counts, it is computed
within a cylinder of radius $r_{200}$. 
The mean (log) richness error is 0.06 dex. 
Richness errors account for Poisson fluctuations in background+cluster
counts and the uncertainty in the mean background counts, as detailed in
previous works (e.g., Andreon \& Congdon 2014; Andreon \& Hurn 2010).
The derivation of these errors assumes a Gaussian approximation, which
is only partially satisfied for these data. For
this reason,
we also list raw galaxy counts in the cluster
and control field directions, and the ratio of the solid angles in which
they are computed. These values are used for 
the mass-richness fit, therefore removing the Gaussian approximation
of the listed $n_{200}$ error.

\section{Results}

\subsection{The mass-richness scaling}

Following previous works, we fit
the data with a linear relation on
log-quantities:
\begin{equation}
\log M_{200} = \alpha+\beta (\log n_{200} -2.0)+\gamma \log\frac{1+z}{1.15}
\end{equation}
allowing for an intrinsic scatter $\sigma_{intr}$, 
noisy mass errors and adopting pivot values of richness and $1+z$
close to the sample mean (the values $2.0$ and $1.15$ in equation 1). 
We adopt weak priors on parameters 
(see Andreon \& Hurn 2010 for details). 
To account for the noisiness of mass errors,
our baseline analysis uses $\nu=6$ to quantify 
that we are 95\% confident that quoted mass errors are correct 
up to a factor of 2 (as mentioned) and  
we anticipate that results are robust to 
the choice of $\nu$. 
Since all clusters in the sample
have a measured richness, there is no selection function (at a
given mass) to account for (see
Andreon \& Berg\'e 2012; Andreon \& Congdon 2014).
We choose not to account for the data structure
(the mass function and the induced Malmquist-Eddington correction
in the astronomical parlance (see Andreon \& Berg\'e 2012 for details)
because we have a limited aim in this work: we acknowledge that
results are derived, and only valid, for clusters drawn from an
X-ray selected sample in a given redshift range\footnote{Note added in proof: 
this statement is overly restrictive, see Andreon (2015b)
for details.}. 

\begin{figure*}
\centerline{
\psfig{figure=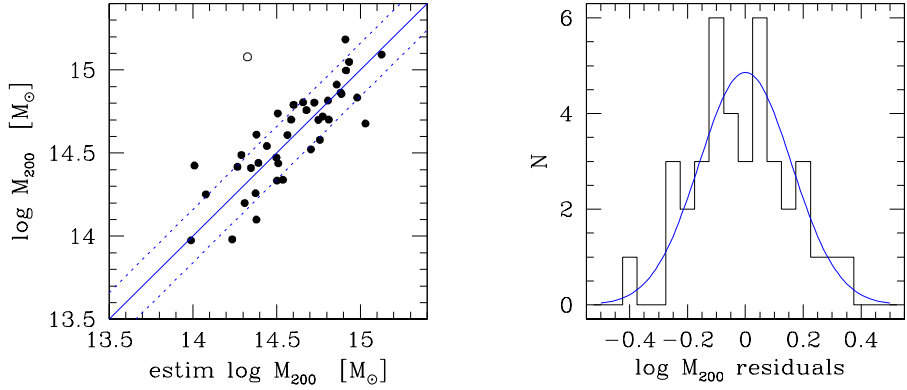,height=5.5truecm,clip=}%
}
\caption[h]{Performances of the richness proxy (i.e., when the 
true $r_{200}$ is unknown). {\it Left-hand panel:} mass-predicted
from richness vs true (caustic) mass. The solid line is
the one-to-one relation (not a fit to the data). The dotted lines show the 
above line plus or minus the data scatter ($0.16$ dex). The open point marks
the outlier Abell 1068 cluster.
{\it Right-hand panel:} Mass residuals (predicted minus true) with superposed
a Gaussian with $\sigma=0.16$ dex.
}
\end{figure*}

The fit to mass, richness, and redshift is shown in Fig.~2. 
Abell 1068 (the open point in Fig.~2) is an outlier.
This cluster is also an outlier of the $L_X$-mass relation
(Rines et al. 2013), and we checked that it is also 
an outlier in the relation between caustic mass and the X-ray
mass given in Piffaretti et al. (2011). We therefore removed this cluster
from the fitting and analysis, although we show it (with a different
symbol) in the figures. For display purposes only, the right-hand panel 
shows a binned version
of the left-hand panel, by combining clusters in bins of eight clusters each,
except the most massive one formed by six clusters.

Fitting the individual values, we found 
that richness scales almost linearly with mass 
(with slope $s=1.30\pm0.10$), with a negligible redshift-dependent term
(slope $-0.1\pm1.0$),
and with a negligible intrinsic scatter. 
More precisely,
\begin{eqnarray}
\log M_{200} &= 14.86\pm0.03+(1.30\pm0.10) (\log n_{200} -2.0) + \nonumber \\
 &\quad (-0.1\pm1.0) \log\frac{1+z}{1.15} .
\end{eqnarray}

The evolution of the richness-mass relation is  poorly determined
because of the reduced redshift baseline ($\Delta z = 0.1$), compared
to the more accurate determination in Andreon \& Congdon (2014) benefiting
from a much larger redshift baseline, $0.15<z<0.55$, 
but using richness and mass measured within a fixed metric aperture. 

The intrinsic scatter is, as mentioned, negligibly small:
the amplitude of the intrinsic
scatter is set by the difference of the data scatter and 
measurement errors. The former is $0.12$ dex, the latter
are $\sim0.08$ dex on both axis, giving little or no room
for an additional intrinsic scatter. In fact, our Bayesian analysis 
found an intrinsic scatter of
$\sigma_{\log M_{200}|n_{200}} < 0.05$ dex with 90\% probability,
consistent and more precise than
the value found using weak-lensing masses in Andreon \& Congdon (2014).
The negligible intrinsic scatter between richness (measured in
cylinders) and mass (measured in spheres) indicates a negligible
effect of large-scale-structure on the richness proxy, unless
it is strongly covariant with cluster elongation along the line of sight. 
The found tight richness-mass scaling is one of the two main 
results of this work.

The negligible scatter of the richness-mass scaling (posterior
mean: $0.02$ dex) scores well
compared to those of best survey-based mass proxies: stellar mass 
($0.08$ dex, Andreon 2012, based however on a reduced cluster
sample), $Y_{sph}$ ($0.09$ dex, Marrone et al. 2012), and
Planck $Y_{SZ}$ ($>0.06$ dex, Planck collaboration 2011).  
The negligible scatter of $n_{200}$ also fares well compared to proxies that requires
follow-up observations, such as 
gas mass ($\sim0.06$ dex, Mahdavi et al. 2013) 
or pseudo-pressure $Y_X$ ($\sim 0.04$ to $\sim 0.10$ dex, Arnaud et al. 2007;
Mahdavi et al. 2013).

We improve upon our previous work (Andreon \& Hurn 2010), based on the 
CIRS sample of Rines et al. (2006), by reducing the uncertainties
of intercept and slope by 30\%, and the uncertainty of intrinsic scatter
by a $>3$ factor as a result of better understood errors and
better determined cluster centers (especially important when 
determining the intrinsic scatter, see also Andreon 2012). 
Unfortunately, the two samples cannot be easily combined
because they adopt slightly different definitions of $M_{200}$.

To summarize, the 39 clusters studied in this work and the 
23 massive clusters at $0.15 < z < 0.55$ in Andreon \& Congdon (2014)
agree in establishing an almost scatterless, 
minimally evolving (if at all), 
mass-richness scaling, in broad agreement with the
53 clusters at lower redshift studied in Andreon \& Hurn (2010).

\begin{figure*}
\centerline{
\psfig{figure=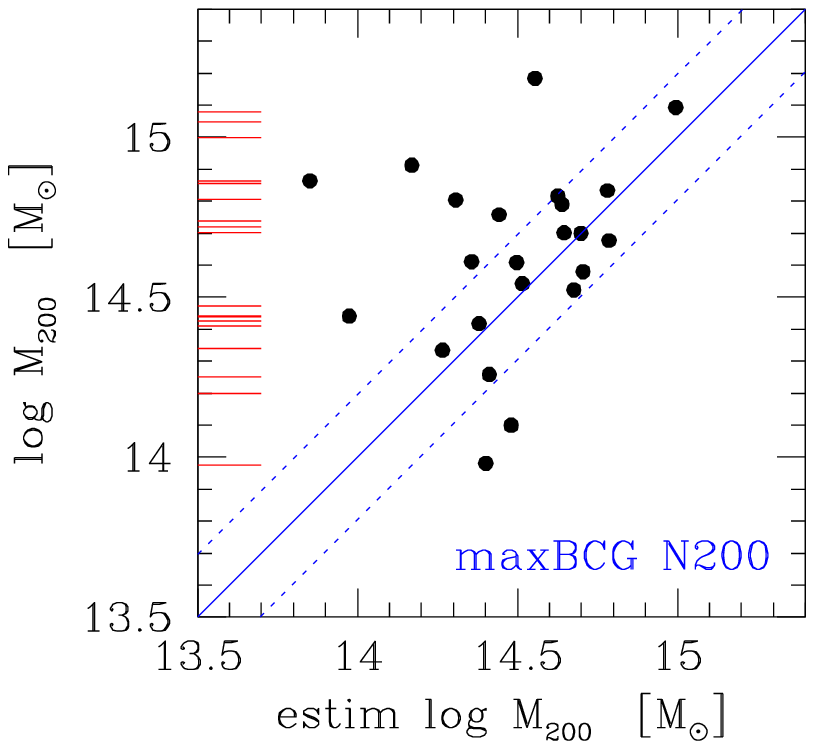,height=5.5truecm,clip=}%
\psfig{figure=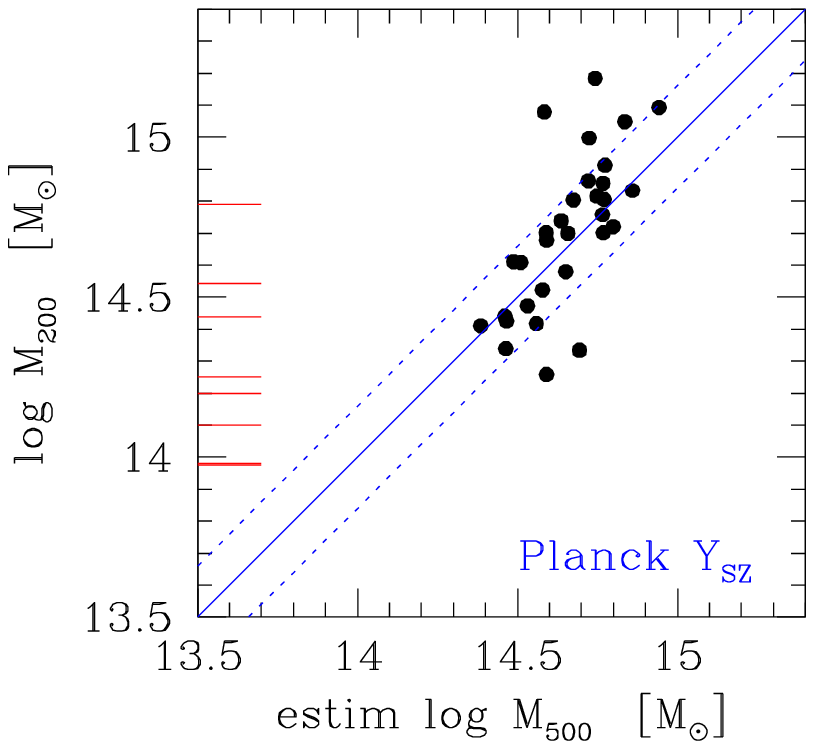,height=5.5truecm,clip=}%
}
\caption[h]{Performances of maxBCG richness (left-hand panel) and of Planck 
$Y_{SZ}$ (right-hand panel) proxies.  The plots compare the 
richness- or $Y_{SZ}$-predicted masses vs the caustic mass for the
very same sample used in Fig.~3. The solid line is
the one-to-one relation (not a fit to the data). The dotted lines show the 
above line plus or minus the claimed mass precision (left-hand panel)
or our richness performances (right-hand panel) for comparison
with Figure 3. 
The (red) ticks on the ordinate of the three panels indicates unmatched  
clusters.
}
\end{figure*}

\subsection{An improved mass proxy estimate}

The tightness of the richness-mass scaling makes
richness an interesting mass proxy. However, as remarked several
times, but not always appreciated, if the observable itself 
requires knowledge of the cluster mass (because knowledge of
$r_{200}$ is needed) such an observable is not ready to be used
as mass proxy until a way to estimate $r_{200}$ is found. 
In this section we make richness an 
effective mass proxy by finding a way to estimate $r_{200}$ that
minimally degrades the performances of the mass proxy.

In Andreon \& Hurn (2010) we measured the cluster richness within a
fixed aperture, the relation between $r_{200}$ and this (aperture) richness
is calibrated with real data and used to infer
the aperture radius, $\widehat{r_{200}}$, in which richness 
should be finally computed (and mass estimated). 
Because of the inherent scatter in the individual radial profiles of galaxy
clusters between aperture radius and $\widehat{r_{200}}$, 
a large amount of scatter is introduced in the mass-proxy scaling.

Kravtsov et al. (2006) proposed a different
approach for the $Y_X$ mass proxy, later also adopted for the $Y_{SZ}$
proxy (Planck Collaboration 2015a) and hereafter for richness: 
$r_{200}$ and the proxy value ($n_{200}$ in our case) are estimated
{\it at the same time} assuming that the proxy-mass relation is
scatterless. In practice, an iterative approach is taken: a radius $r$ is
taken (1.4 Mpc in our case), $n(<r)$ estimated, then $r$ is updated
to the value appropriate for the derived richness if the mass-richness
were scatterless (i.e., using eq. 2, and noting that $r_{200}=M^{1/3}_{200}$
apart from obvious coefficients) and then the process is iterated
until convergence. In our case, judging from comparison with
results obtained after ten iterations, convergence is achieved at
the second or third iteration. We therefore adopt the fifth iteration
as the final (but any other iteration would give equivalent results). 
These values of richness and masses, derived without knowledge of
the true $r_{200}$, are listed in Table 1.

The left-hand panel of Figure~3 shows the performance of the richness
proxy when ready to be used for cluster samples of unknown mass (i.e.
without knowledge of the true $r_{200}$): the richness in
the estimated $r_{200}$, converted in mass using eq. 2, i.e., the
richness-predicted mass, is plotted versus the
caustic mass. The scatter between the richness-predicted mass
and mass is $0.16\pm0.02$ dex (error computed by bootstrapping), 
also shown in the right-hand panel of Figure 3.
Of this, $0.08$ dex is due to the noisiness of the caustic mass estimate
(i.e., caustic errors).  

Our quoted error of the richness-based mass is total, in contrast to 
what is quoted for
other mass estimates. For example, it includes 
the scatter around the relation and errors on the observable.
Our choice of quoting all known
error terms in the error budgets is not widespread: for example the error quoted
by the Planck team for the $Y_{SZ}$-based masses 
does not include the terms
above (sec 5.3 of Planck Collaboration 2015a). Finally,
and in contrast to some other claims about the error of other
proxies, the scatter we derived above is not computed assuming we know
the true $r_{200}$. If we had to follow these works, we should
claim a $\sim 0.02$ dex error, not the quoted $0.16$ dex error.

As mentioned, we removed one single cluster pair (Abell
750/MS0906.5+1110) out of the starting sample (about 40 clusters) 
because one is projected
on the top of the other and with $\Delta v \sim 3000$ km/s,
leading to an erreneous richness (and lensing mass). This
situation is very rare,
no other cluster has been removed from the sample 
for richness-based reasons, and the other two removed
clusters (Abell 1068 and Abell 2055) have unproblematic richnesses but
questionable caustic masses. If this sample is representative of clusters in the
Universe, we then expect that richness-based masses will be very wrong for
only about 2\% of the clusters.

The above predicted mass uses accurate cluster centers.
Since we also want to exploit richness as mass proxy for clusters 
without accurate
centers, we now check the impact of lacking accurate centers. We
therefore implement a way to estimate centers.
We iteratively derive the cluster center only using galaxy counts:
we consider an aperture of 1.0 Mpc radius, we compute the median 
right ascension
and declination, we consider this the  new center, and we iterate ten
times. We start the center computation 3 arcmin, about
0.5 Mpc at the median redshift, away from 
the true cluster center to avoid the advantage of 
starting from the optimal center.
We take the latest computed centers
as new cluster centers (but results would be unchanged using, e.g.,
the fifth iteration), and we solve for radius and mass as before. 
We found an identical scatter between the richness-predicted mass
and mass, as expected because the center error is negligible 
compared to $r_{200}$.

Our current richness scores best among all mass proxies 
based on survey data, as far as we are aware of, as we now illustrate. 

First, the left-hand 
panel of Figure~4 shows the performance of the maxBCG richness 
using the mass calibration in Rozo et al. (2009) via
plotting the richness-predicted mass vs caustic mass for the very same 
HECS sample shown in Fig.~3.
While using similar photometry (a previous SDSS release), 
these authors adopt different recipes for
counting the galaxies (notably the adopted aperture)
and a different (indirect) approach
to mass calibration. Clusters are matched with a 2
arcmin maximum radius and our results are robust to 
adopted matching aperture.
Clusters not found in the maxBCG catalog (about 40\%) 
are shown as ticks on the ordinate. For the remaining 60\%, we found
a scatter of $0.26$ dex (consistent with, but larger than, 
the advertised 
value). Our mass proxy
performs better, both in terms of scatter and 
in terms of completeness. Adopting the more recent
calibration by Rozo et al. (2014) does not change our results.
By taking the richest maxBCG cluster with consistent redshift
and within a large matching radius does not change our results either.
The improved
maxBCG richness by Rykoff et al. (2012) shows a reduced scatter (although
larger than advertised) and a large mass bias, the latter anticipated
by the authors because of their preliminary calibration. By adopting
a direct approach to mass calibration, our richness is not 
badly affected by systematics. More recent catalogs
based on SDSS lack a mass calibration. Considering
them would require us to calibrate them first, which is
beyond the purpose of this paper.

Second, the right-hand panel of Figure 4 illustrates the performances
of $Y_{SZ}$-based
masses, taken from Planck collaboration (2015a) via
plotting the $Y_{SZ}$-predicted mass vs caustic mass.
Clusters are matched with a maximal radius of 3 arcmin (equal to about
twice the 68\% Planck position error).
Unmatched clusters (eight cases) 
are marked with red tick on the figure ordinate. 
Some of them are of low mass and are therefore likely absent in the 
Planck catalog because they are undetected. However, some of them are somewhat 
massive. The scatter of the remaining 31 $Y_{SZ}$-based masses  
is comparable to richness-based masses ($0.18$ vs $0.16$ dex), 
but perhaps 
better described by a combination of a slightly narrower distribution (with
$\sigma\sim0.14$ dex) plus some ($15$\%) outliers. 
Enlarging the matching
radius does not help to increase the $Y_{SZ}$ mass performances because
the newly matched objects have large scatter with mass. 
Decreasing the matching radius is of no help either
because it decreases the number
of matches without reducing the scatter.
By restricting our attention to clusters
in the narrow redshift range studied in this work 
($0.10<z<0.22$) and by
computing the {\it minimal} scatter (by allowing
a free mass bias),  we minimize known redshift- (Andreon 2014) 
and bias- (von der Linden et al. 2014, Planck collaboration 2015b) 
effects of $Y_{SZ}$-based
masses.
To summarize our comparison with $Y_{SZ}$-based masses, proxy availability 
and a low intrinsic scatter with mass  
are essential to evaluate the quality of
a mass proxy.The 54 sec exposures taken at a 2m telescope in a mediocre site
on Earth (the SDSS) competes well with a
satellite at the L2 point from this point of view in the studied
redshift range\footnote{One may argue that we favored
optical richness by removing  
the cluster pair MS0906.5+1110/Abell 750 from the sample. However, $Y_{SZ}$-based
masses suffer of the same alignment problem because the $Y_{SZ}$
observable lacks almost any redshift sensitivity.}.

Third, our score of a $0.16$ dex scatter is better than all 25 methods
considered in the mass reconstruction project (Old et al. 2015), most 
of which requires spectroscopy, i.e., are observationally more 
demanding. In particular, in these simulations 
richness-based masses outperform
caustic masses. If the same holds true
in our Universe, our work would be measuring the
scatter of a precise mass estimate (richness) via comparison
to a noisy estimate (caustic masses)!
Note however that such a simulation-based analysis
should be considered with caution
because results depends on the way clusters are populated 
by galaxies in the simulations, and because nothing guarantee that 
the best mass proxy in 
simulations is also best in our Universe.

We recal that a tight mass-proxy
scaling is certainly useful, but scalings that involve 
observationally 
parsimonious observables are even better, and richness is both 
parsimonious and display a tight relation with mass.
Indeed, richness has already allowed us
to derive the mass of three among the most distant clusters and groups known 
(Strazzullo et al. 2014; Andreon et al. 2014; Webb et al. 2015), which still miss $Y_X$-
or $Y_{SZ}$-based
masses because they are observationally expensive to acquire.

Cluster counts are
exponentially sensitive to the properties of dark energy and
to the square of the scatter between proxy and mass.
Our factor two improvement over Andreon \& Hurn (2010)
and maxBCG,
obtained because the current estimate benefits 
from a radius measured very close to the
true $r_{200}$ and tailored to each cluster,
corresponds to a four times lower 
dilution of cosmological information contained in number counts.
This improvement is our second main result.

\section{Conclusions}

We studied 39 X-ray selected 
clusters at $0.1<z<0.22$ with accurate caustic masses
(20\% errors) from the HECS catalog. We
derived richness within $r_{200}$ using SDSS photometry
and we found an extremely tight, almost scatterless, richness-mass 
scaling, at least as tight as those derived for other mass proxies,
and tighter than quoted in previous works. 
The current sample, adjoinining 76 more
clusters analyzed in previous works, establishes
an almost scatterless, minimally evolving (if at all), 
mass-richness scaling in the redshift range $0.03<z<0.55$.
By assuming
that the true relation is scatterless, we rederive cluster richness
and $r_{200}$ at the same time only using SDSS photometry. We found
that the newly derived richness is a mass proxy with $0.16$ dex scatter,
half of which is not
intrinsic to the proxy, but related to the noisiness of the 
caustic masses used for test richness performances.

To summarize, 
an ideal mass proxy should be
characterized by a low intrinsic scatter, be observationally parsimonious to
obtain, relatively insensitive to the cluster 
dynamical state, and easy to compute. Richness seems to be as such. For this
reason we are now computing richness-predicted masses of many $z<0.22$ clusters
in the SDSS footprint, with priorities set by the community input.

\begin{acknowledgements}
This work benefits of discussion with caustic specialists, Antonaldo Diaferio,
Ana Laura Serra, and Jacob Svensmark, and with Andrea Biviano.
\end{acknowledgements}

{}

\end{document}